\documentstyle[amssymb,aps,psfig,epsf]{revtex}
\begin{document}
\title{Long-range thermoelectric effects in mesoscopic superconductor-normal metal
structures. \\
}
\author{A.F. Volkov$^{1,2}$ and V. V. Pavlovskii$^{2}$}
\address{$^{(1)}$Theoretische Physik III,\\
Ruhr-Universit\"{a}t Bochum, D-44780 Bochum, Germany\\
$^{(2)}$Institute of Radioengineering and Electronics of the Russian Academy%
\\
of Sciences, 103907 Moscow, Russia }
\maketitle

\begin{abstract}
We consider a mesoscopic \ four-terminal superconductor/normal metal (S/N)
structure in the presence of a temperature gradient along the N wire. A
thermoemf arises in this system even in the absence of the thermoelectric
quasiparticle current if the phase difference between the superconductors is
not zero. We show that the thermoemf is not small in the case of a
negligible Josephson coupling between two superconductors. It is also shown
that the thermoelectric voltage has two maxima: one at a low temperature and
another at a temperature close to the critical temperature. The obtained
temperature dependence of the thermoemf describes qualitatively experimental
data.
\end{abstract}

\section{Introduction}

Transport phenomena in mesoscopic superconductor-normal metal (S-N)
structures have attracted a great interest in last years (see for example
reviews \cite{Beenakker,LambertR} and references therein). Due to the
proximity effect (PE) such properties of the normal metal N as the
density-of-states etc are changed, and therefore transport properties of S-N
structures are changed as well.\ For example, the resistance of a normal
wire a part of which is a segment of a superconducting loop is a periodic
function of a magnetic flux $\Phi $ threading this loop. The period of the
oscillations is the magnetic flux quantum $\Phi _{0}=hc/2e.$ The amplitude
of the resistance oscillations $\delta R_{m}$ depends on temperature $T$ in
a non-monotonic way reaching a maximum at $T_{m}=c_{1}\epsilon _{L}$, where $%
c_{1}$ is a numerical factor and $\epsilon _{L}=D_{N}/L^{2}$ is the Thouless
energy \cite{Nazarov,Pannetier}. A lot of publications are devoted to the
study of the electric conductance of S-N mesoscopic structures. Much less
attention was paid to the study of thermoelectric phenomena in such
structures. However the situation has changed recently. A number of papers
has been published in which the results of both experimental \cite
{Chandrasekhar98,Chandrasekhar02,Petrashov} and theoretical \cite
{LambertT,SV,KPV,Falko,Vinokur,H} studies are presented. It has been
established experimentally that the thermoemf $V_{th}$ in a S-N structure
with a superconducting loop also is an oscillating function of the magnetic
flux in the loop and the amplitude of oscillations depends on temperature in
a non-monotonic way \cite{Chandrasekhar98,Chandrasekhar02,Petrashov}.

In Refs. \cite{Falko,Vinokur,Chandrasekhar05} the thermoconductance in S-N
mesoscopic structures was calculated and measured. These calculations
generalized the results for the thermoconductivity in superconductors in the
intermediate state obtained by Andreev a long time ago \cite{Andreev}. The
thermoelectric effects in superconductors caused by the thermoelectric
component $j_{th}=\eta \nabla T$\ in the quasiparticle current $j_{qp}$ were
considered in many papers starting from the Ginzburg paper \cite{Ginzburg}
(for more references see \cite{Gurevich}). These effects in a short bridge
between two superconductors were studied recently in Ref.\cite{Gurevich}.
The authors established that due to a charge imbalance the thermoemf in the
superconducting state may be comparable with that in the normal state. The
thermoemf $V_{th}$ induced in mesoscopic S-N structures with normal
reservoirs kept at different temperatures was theoretically studied in Refs.
\cite{LambertT,SV,KPV,H}. Using a scattering matrix approach, Claughton and
Lambert \cite{LambertT} studied the dependence of the thermoelectric voltage
on the phase difference $\varphi $ between the superconductors S taking into
account the thermoelectric current $j_{th}=\eta \nabla T$. They showed that
this dependence is periodic.

Unlike Ref.\cite{LambertT} in Refs. \cite{SV,KPV} the thermoelectric current 
$j_{th}$ was completely neglected because the thermoelectric coefficient $%
\eta $ contains a small parameter: $\eta \propto T/\epsilon _{F}$, where $%
\epsilon _{F}$ is the Fermi energy. It was shown that even in the absence of
the thermoelectric current $j_{th}$ a temperature gradient in the N wire
leads to the build-up of a thermoelectric voltage $V_{th}$ if the N wire is
in contact with two superconductors and the phase difference $\varphi $
between the superconductors is not zero. The voltage $V_{th}$ in the S-N
structure may be of the order $V_{th}\approx c_{2}\delta T/e$, where $\delta
T=T_{r}-T_{l}$, $T_{r,l}$ is the temperature of the right (left) normal
reservoir (see Fig.1), $c_{2}$ is a numerical factor (in the case under
consideration $c_{2}\approx 0.1$). Therefore the thermoemf $\ V_{th}$ below $%
T_{c}$ is much larger than the ordinary thermoemf \ $V_{ord}$ ($V_{ord}\sim
(T/\epsilon _{F})\delta T/e$) above the critical temperature $T_{c}$. This
effect can be called a giant thermoelectric effect in S-N mesoscopic
structures. It was shown in Refs. \cite{SV,KPV} that the voltage $V_{th}$
oscillates with increasing the phase difference: $V_{th}\propto \sin \varphi 
$, that is, the function $V_{th}(\varphi )$ is shifted by $\pi /2$ with
respect to the phase dependence of the resistance variation $\delta R_{N}$
of the normal wire: $\delta R_{N}(\varphi )\propto \cos \varphi $. The
amplitude of these oscillations\ $V_{th}(\pi /2)$ is a non-monotonic
function of temperature with a maximum at a temperature $T_{m}$ of the order
of the Thoulless energy $\epsilon _{L}.$ These results are obtained in Refs. 
\cite{SV,KPV} on the basis of microscopic equations for the quasiclassical
Green's functions. It was assumed that the S-N interface transmittance is
low (due to a mismatch of the Fermi surfaces or to the presence of a
potential barrier) and therefore the PE is weak. In this case the problem
can be solved analytically.

Another limit of perfect S-N interfaces was considered in Ref.\cite{H} by
using the same microscopic equations. Solutions for these (kinetic and
Usadel) equations were found in Ref.\cite{H} numerically and the obtained
results are similar to those found in Refs. \cite{SV,KPV}. In the
theoretical publications \cite{SV,KPV,H} the physics of this giant
thermoelectric effect is explained in terms of the temperature-dependent
Josephson current which, in the presence of a temperature gradient, has
different values at different S-N interfaces. In order to compensate this
difference, a quasiparticle current driven by the voltage $V_{th}$ arises in
the system. An approximate formula for the thermoemf was presented in Ref. 
\cite{H}, where in the main approximation the voltage $V_{th}$\ is expressed
in terms of the Josephson critical current $I_{c}(T)$. A correction to this
expression for the voltage $V_{th}$\ is small. Although in some limiting
cases the representation of $V_{th}$ through $I_{c}(T)$ is possible (the
authors assumed that the order parameter $\Delta $\ is much larger than the
Thouless energy $\epsilon _{L}$), in a general case this can not be done.
The point is that the Josephson current is a thermodynamical quantity (it
can be presented as a derivative of a free energy with respect to the phase
difference), whereas the voltage $V_{th}$ is not. To make this point quite
clear, one can consider a limiting case of temperatures $T$ larger than $%
\epsilon _{L}$ (to be more exact, the ratio $2\pi T/\epsilon _{L}$ should be
much larger than 1). In this case the Josephson current is exponentially
small and the Josephson coupling is negligible. However, generally speaking,
the voltage $V_{th}$ is not exponentially small. Its value is determined by
the ratio of other parameters: $\Delta $ and $\epsilon _{L}$. The aim of
this paper is to consider the case when the Josephson current is small, but
the thermoelectric effect is not small in comparison with its maximal value.
We will show that if the ratio $2\pi T/\epsilon _{L}$ is large, the
Josephson coupling between superconductors is almost negligible, but the
thermoemf $V_{th}$ is not small and can be measured. In this case one can
say about a long-range thermoelectric effects.

\section{\protect\bigskip Model and basic assumptions}

As in Refs.\cite{SV,KPV,H}, we consider a system shown schematically in the
inset of Fig. 3. A normal wire N, or a thin film with a width narrower than
the coherence length\bigskip\ $\xi _{T}=\sqrt{D_{N}/2\pi T},$ connects two
normal reservoirs N$_{l,r}$. The left reservoir N$_{l}$ is kept at a
temperature $T$ and the right reservoir N$_{r}$ has a temperature $T+\delta
T $. We assume that the S-N interface resistance $R_{b}$ is larger than the
resistance $R_{L}$ of the N wire so that the ratio $R_{L}/R_{b}$ is a small
parameter. In this case the problem allows an analytic solution because
there is a small parameter, the amplitude of the condensate function $\mid
F^{R(A)}\mid ,$ where $F^{R(A)}$ is the retarded (advanced)\ quasiclassical
Green's function induced in the N wire due to the PE. Therefore the PE is
assumed to be weak and the characteristics of the N wire deviate only
slightly from those in the normal state.

The thermoemf $V_{th}$ arising in the system at $\delta T\neq 0$ is an
integral over energies $\epsilon $ from a distribution function $%
f_{-}(\epsilon )$ \cite{SV,KPV,H}. This distribution function called a
transverse part of the distribution function in Ref. \cite{SS} and denoted
by $f_{1}$ in Ref. \cite{LO} is a difference between electron- and hole-like
excitations: $f_{-}=n_{\uparrow }-p_{\downarrow }$(see, for example, \cite
{AV} and \cite{Kopnin}). This function determines the voltage and is related
to a so-called charge-imbalance \cite{Tinkham}. Another distribution
function $f_{+}=1-(n_{\uparrow }+p_{\downarrow })$\ determines, for example,
the supercurrent and the order parameter (in the superconductor). These
distribution functions obey kinetic equations \cite{SS,LO,AV,Kopnin}. The
kinetic equations were applied to the study of transport properties of
mesoscopic S/N structures \cite{VZK}{\bf .} The function $f_{-}$ obeys a
kinetic equation, which being written in notations of Ref. \cite{SV,KPV} has
the form

\begin{equation}
{\ \ M}_{-}{\ \partial }_{x}f_{-}(x){\ +J}_{S}f_{+}(x)-{\ J}_{an}{\ \partial 
}_{x}f_{+}(x){\ =J}  \label{f_m}
\end{equation}
where all the coefficients are expressed \ in terms of the retarded
(advanced) Green's functions: $\widehat{G}^{R(A)}=G^{R(A)}\hat{\sigma}_{z}+%
\widehat{F}^{R(A)};$ $\ M_{-}=(1-G^{R}G^{A}-(\widehat{F}^{R}\widehat{F}%
^{A})_{1})/2;$ $J_{an}=(\widehat{F}^{R}\widehat{F}^{A})_{z}/2,\ $is an
anomalous current, and the notation $(\widehat{F}^{R}\widehat{F}^{A})_{z}$
means $(\widehat{F}^{R}\widehat{F}^{A})_{z}=Tr\{\widehat{\sigma }_{z}%
\widehat{F}^{R}\widehat{F}^{A}\}/2.$ The function $J_{S}$ determines the
Josephson current and can be expressed through the values of $\widehat{F}%
^{R(A)}$ at the S/N interface

\begin{equation}
J_{s}=-(i/(4R_{b}\sigma ))Tr\{\widehat{\sigma }_{z}(\widehat{F}^{R}\widehat{F%
}_{s}^{R}-\widehat{F}^{A}\widehat{F}_{s}^{A})\}
\end{equation}
The Green's functions $\widehat{F}_{S}^{R(A)}$ in the superconductors S are
not disturbed by the PE due to a low S/N interface transmittance, and
therefore in the right superconductor, they are equal to

\begin{equation}
\widehat{F}_{S}^{R(A)}=[i\hat{\sigma}_{y}\cos (\varphi /2)+i\hat{\sigma}%
_{x}\sin (\varphi /2)]\Delta /\xi _{\epsilon }^{R(A)}  \label{F_S}
\end{equation}
where $\xi _{\epsilon }^{R(A)}=\sqrt{(\epsilon \pm i\gamma )^{2}-\Delta ^{2}}
$; a phenomenological parameter $\gamma $ describes a damping in the
superconductors; the matrices $\widehat{F}_{S}^{R(A)}$ in the left
superconductor have the same form if one makes a replacement: $\varphi
\Longrightarrow -\varphi $ . The coefficient $M_{-}$ is also expressed
through the retarded (advanced) Green's functions $F^{R(A)}$ in the N wire,
but the corrections due to the weak PE are small and approximately we have $%
M_{-}\approx 1$. The partial total current $J$ is constant over each
separate piece of the N wire. Since the ''currents''$J_{S}$ and $J_{an}$ are
proportional to small parameters $\mid F^{R(A)}\mid ^{2},$ the distribution
function $f_{+}$ (longitudinal in terms of Ref.\cite{SS}) may be taken in
zero order approximation, that is, equal to its value in the absence of the
PE. It has the form

\begin{equation}
f_{+}\cong \lbrack \delta f_{eq}/2](1+x/L_{N})  \label{f_p}
\end{equation}
where $\delta f_{eq}=[\tanh (\epsilon /2(T+\delta T))-\tanh (\epsilon /2T)].$
Eq.(\ref{f_m}) should be solved with the boundary conditions at the points $%
x=\pm L_{N}:$ $f_{-}(\pm L_{N})=F_{-}(V_{r,l})$, where $F_{-}(V)=[\tanh
((\epsilon +eV)/2T)-\tanh ((\epsilon -eV)/2T)]/2$. In addition, one has to
use the boundary conditions at the S/N interfaces (see \cite{SV,KPV} ). The
voltages $V_{\pm }\equiv (V_{r}\pm V_{l})/2$ are found from the condition of
the absence of the total current through the N reservoirs (the condition of
a disconnected circuit). They are equal to \cite{KPV,Foot}

\begin{equation}
eV_{+}\cong -\delta T(L_{1}/L_{N})\int d\epsilon (\epsilon \beta
)g_{z+}(\epsilon ,L_{1})f_{eq}^{\prime }/\int d\epsilon ((\upsilon
_{S}+g_{1+}(\epsilon ,L_{1}))f_{eq}^{\prime });  \label{V_p}
\end{equation}
\begin{equation}
2eV_{-}\cong r_{S}\delta T\int d\epsilon (\epsilon \beta )g_{z-}(\epsilon
,L_{1})f_{eq}^{\prime }  \label{V_m}
\end{equation}
where $g_{z\pm }=(1/4)[(\widehat{F}^{R}\mp \widehat{F}^{A})(\widehat{F}%
_{S}^{R}\pm \widehat{F}_{S}^{R})]_{z},$ $g_{1+}=(1/4)[(\widehat{F}^{R}+%
\widehat{F}^{A})(\widehat{F}_{S}^{R}+\widehat{F}_{S}^{A})]_{1}$, $%
f_{eq}^{\prime }=\cosh ^{-2}(\epsilon /2T)$, and $\upsilon _{S}=%
%TCIMACRO{\func{Re}}%
%BeginExpansion
\mathop{\rm Re}%
%EndExpansion
G_{S}^{R}(\epsilon )$ is the density-of-states of a BCS superconductor, $%
r_{S}=R_{S}/R_{b}$ is a small parameter, $R_{b}$ is the S/N boundary
resistance, $R_{S}=L_{1}/\sigma $ is the resistance of a piece of the N wire
in the normal state. The denominator in Eq.(\ref{V_p}) is proportional to
the conductance of the S/N interface \cite{VZK,Hekking,VPhysB}. The first
term in the integrand of the denominator describes the conductance due to
quasiparticles above the gap $\Delta ,$\ whereas the second term describes
the subgap conductance ($|\epsilon |<\Delta $). Therefore the integrand is
not zero at all energies.

Thus the voltages $V_{\pm }$ are expressed in terms of the Green's functions 
$F^{R(A)}$ which can be found from the linearized Usadel equation. Discuss
now some general properties of the expressions (\ref{V_p}) and (\ref{V_m}).
The condensate functions $F^{R(A)}$ induced in the N wire have the same
matrix structure as the functions $\widehat{F}_{S}^{R(A)}$ (see Eq.(\ref{F_S}%
))

\begin{equation}
\widehat{F}^{R}=i\hat{\sigma}_{y}\cos (\varphi /2)F_{y}^{R}+i\hat{\sigma}%
_{x}\sin (\varphi /2)F_{x}^{R}  \label{F}
\end{equation}
Taking into account that $F^{R}=-(F^{A})^{\ast }$, the expressions for $%
g_{z\pm }$ and for $g_{1+}$ can be represented in the form

\begin{eqnarray}
g_{z+} &=&\frac{1}{2}%
%TCIMACRO{\func{Re}}%
%BeginExpansion
\mathop{\rm Re}%
%EndExpansion
(F_{y}-F_{x})%
%TCIMACRO{\func{Im}}%
%BeginExpansion
\mathop{\rm Im}%
%EndExpansion
F_{S}\sin \varphi ;\text{ }g_{z-}=\frac{1}{2}%
%TCIMACRO{\func{Im}}%
%BeginExpansion
\mathop{\rm Im}%
%EndExpansion
(F_{y}-F_{x})%
%TCIMACRO{\func{Re}}%
%BeginExpansion
\mathop{\rm Re}%
%EndExpansion
F_{S}\sin \varphi  \label{g_pm} \\
g_{1+} &=&\frac{1}{2}%
%TCIMACRO{\func{Im}}%
%BeginExpansion
\mathop{\rm Im}%
%EndExpansion
[(F_{y}+F_{x})+(F_{y}-F_{x})\cos \varphi ]%
%TCIMACRO{\func{Im}}%
%BeginExpansion
\mathop{\rm Im}%
%EndExpansion
F_{S}
\end{eqnarray}
Here we dropped the index $R$: $F_{y}^{R}=F_{y}.$ Note also that $%
%TCIMACRO{\func{Im}}%
%BeginExpansion
\mathop{\rm Im}%
%EndExpansion
F_{S}=-\Delta /\mid \xi (\epsilon )\mid $ at $\mid \epsilon \mid \leq \Delta 
$ and $%
%TCIMACRO{\func{Im}}%
%BeginExpansion
\mathop{\rm Im}%
%EndExpansion
F_{S}=0\mid $ at $\mid \epsilon \mid \geq \Delta $, whereas $%
%TCIMACRO{\func{Re}}%
%BeginExpansion
\mathop{\rm Re}%
%EndExpansion
F_{S}=\Delta /\xi (\epsilon )$ at $\mid \epsilon \mid \geq \Delta $ and $%
%TCIMACRO{\func{Re}}%
%BeginExpansion
\mathop{\rm Re}%
%EndExpansion
F_{S}=0$ at $\mid \epsilon \mid \leq \Delta .$ Therefore at low temperatures
($T<<\Delta $) the voltage $V_{-}$ is much smaller than the voltage $V_{+}$.
It is useful to compare Eqs.(\ref{g_pm}) and the expression for $J_{S}$ that
can be represented as 
\begin{equation}
J_{S}=(r_{S}/2L_{1})%
%TCIMACRO{\func{Im}}%
%BeginExpansion
\mathop{\rm Im}%
%EndExpansion
[(F_{y}-F_{x})F_{S}]\sin \varphi  \label{J_S}
\end{equation}

\begin{figure}[tbp]
\epsfysize= 8cm \vspace{0.2cm}
\centerline{\epsfbox{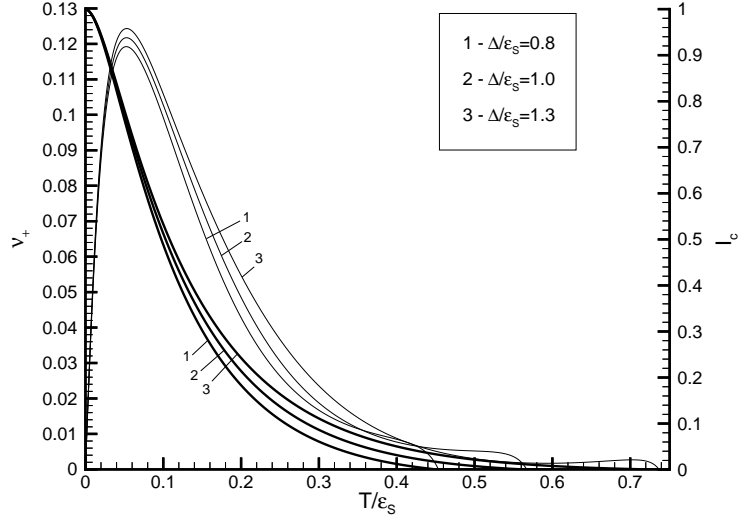}} \vspace{0.2cm}
\large{\caption{Temperature dependence of the normalized thermoemf (the solid lines, the left scale) $v_{+}=e(V_{r}+V_{l})/2\delta T$ and the normalized critical current
(the dashed lines, the right scale) $I_{c}(T)/I_{c}(0)$ for different values of the ratio $%
\Delta /\epsilon _{S}$ and $r_{S}=0.5$; where $\epsilon _{S}=D/L_{1}^{2}$
and $\Delta $ is the energy gap at zero temperature.Results are shown for $%
L_{1}/L_{2}=\sqrt{3}$; $L_{1}/L_{SN}\equiv L_{1}/(L_{N}-L_{1})=\sqrt{0.3}$; $%
\gamma =0.1\epsilon _{S}$.}}

\end{figure}

\begin{figure}[tbp]
\epsfysize= 8cm \vspace{0.2cm}
\centerline{\epsfbox{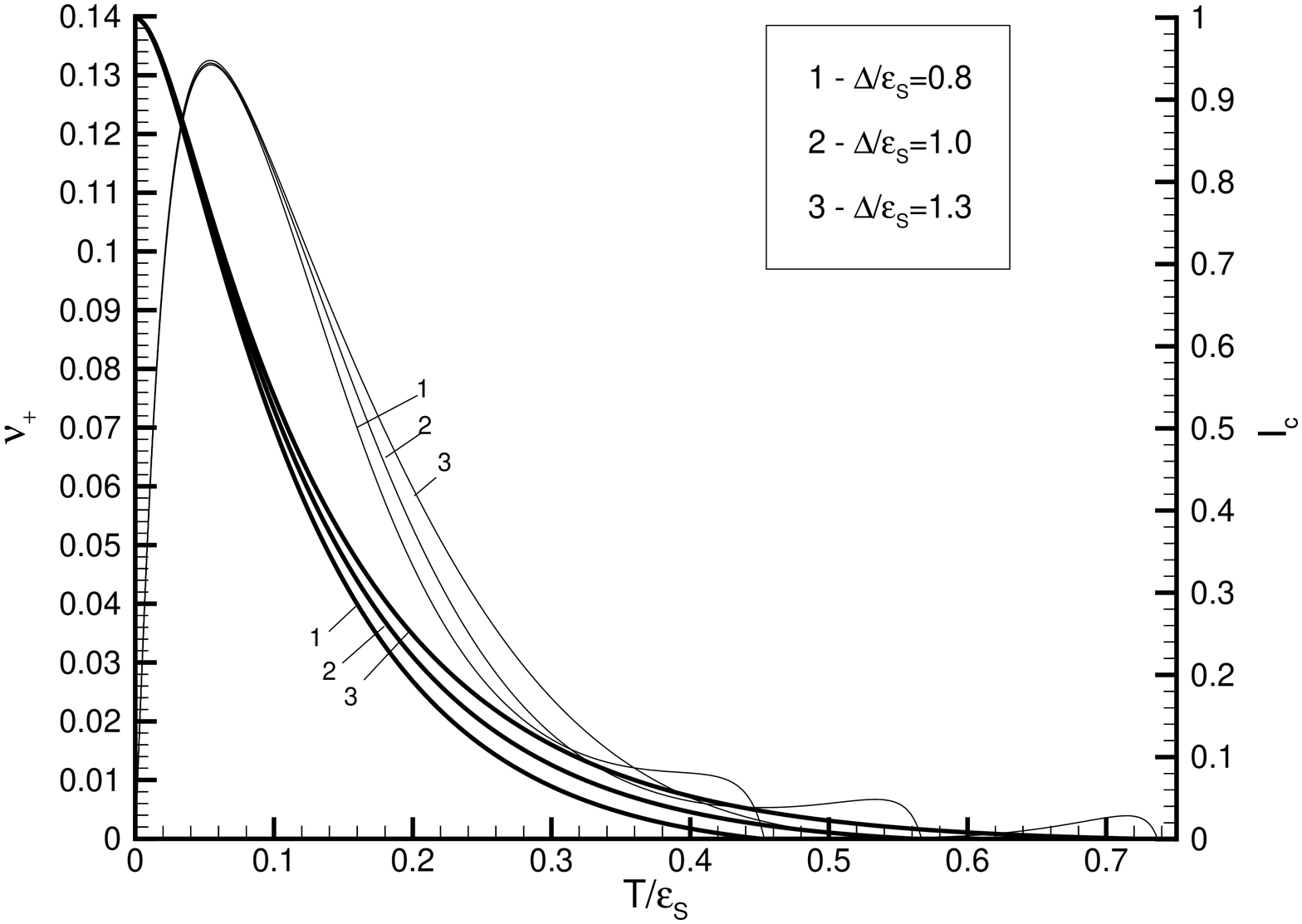}} \vspace{0.2cm}
\large{\caption{The same dependencies as in Fig.1 for
$\gamma =0.03\epsilon _{S}$.}}
\end{figure}

The critical current is an integral over all energies from the function $%
J_{S}(\epsilon )$ which can be transformed into a sum over the Matsubara
frequencies $\omega =\pi T(2n+1)$

\begin{equation}
I_{c}=\frac{\sigma r_{S}}{4L_{1}}\int d\epsilon 
%TCIMACRO{\func{Im}}%
%BeginExpansion
\mathop{\rm Im}%
%EndExpansion
[(F_{y}-F_{x})F_{S}]\tanh (\epsilon /2T)=\frac{\pi T\sigma r_{S}}{4L_{1}}%
\sum_{\omega }[(F_{y}-F_{x})F_{S}]_{\epsilon =i\omega }  \label{I_c}
\end{equation}

We will see that at large ratio $T/\epsilon _{L},$ the difference $%
(F_{y}-F_{x})\propto \exp (-4\sqrt{2\omega /\epsilon _{L}})$ is
exponentially small and therefore the critical current $I_{c}$ is very small
(here we take $\epsilon _{L}=D/L^{2}$ with $L=L_{1}=L_{2}$). On the other
hand, Eq.(\ref{g_pm}) for $g_{z+}$, for example, can not be represented as a
sum over the Matsubara frequencies. The point is that the critical current $%
I_{c}$\ can be written as an integral over energies from the product of the
retarded (or advanced) Green's functions only: $I_{c}\sim 
%TCIMACRO{\func{Im}}%
%BeginExpansion
\mathop{\rm Im}%
%EndExpansion
\int d\epsilon \lbrack (F_{y}-F_{x})F_{S}]^{R}\tanh (\epsilon /2T).$\
Therefore one can enclose the contour of integration in the upper half-plane
of $\epsilon $, where the retarded functions are analytic functions, and
calculate the sum over the poles (the Matsubara frequences) of the function $%
\tanh (\epsilon /2T).$\ The functions $g_{z\pm }$\ contain the products of
the type $(F_{y}-F_{x})^{R}F_{S}^{A}$\ (an anomalous function in terminology
of Ref.\cite{GE}), which are not analytic functions both in the upper and
lower half-plane of $\epsilon .$ Thus the integral can not be reduced to the
sum over the Matsubara frequencies. The importance of the anomalous terms
was emphasized by Gor'kov and Eliashberg \cite{GE} who showed that due to
these terms the generalization of the Ginzburg-Landau equation to a
nonstationary case, generally speaking, is not possible. In the case of
mesoscopic S/N structures these terms lead to long-range effects \cite
{VTakan}.

Thus in a general case the statement about the smallness of $V_{\pm }$ is
not valid. Only if the condition $T,\epsilon _{L}<<\Delta $ is fulfilled,
one can regard $F_{S}$ as a constant (the integrand in Eq.(\ref{g_pm})
converges over energies of the order of $T$) and represent Eq.(\ref{V_m}) as
a sum over the Matsubara frequencies. In this case the voltages $V_{\pm }$
are also exponentially small. However if the Thouless energy $\epsilon _{L}$
is not small in comparison with $\Delta $ (for example, in \cite{Petrashov} $%
\epsilon _{S}\approx $ \ $2.8$ $K$ and $\Delta \approx 2.28$ $K$), one can
not neglect the dependence of $F_{S}$ on the energy $\epsilon $ and
represent the integral in the form of the sum over the Matsubara
frequencies. In this case the critical current $I_{c}$ may be exponentially
small ($I_{c}\propto \exp (-\sqrt{2\pi T/\epsilon _{L}})$), whereas the
voltages $V_{\pm }$ are not small.

In order to calculate the voltages $V_{\pm }$ explicitly, we need to find
the retarded (advanced) Green's function $F^{R(A)}$. In the considered limit
of the weak PE, these functions are easily found from a solution for the
linearized Usadel equation

\begin{equation}
\partial ^{2}\widehat{F}^{R(A)}/\partial x^{2}-(\kappa ^{R(A)})^{2}\widehat{F%
}^{R(A)}=0  \label{Us_Eq}
\end{equation}
where $(\kappa ^{R(A)})^{2}=\mp 2i\epsilon /D_{N}$. We write out here the
solutions for the functions $F_{x,y}$ (we again dropped the indices $R(A)$)

\begin{equation}
F_{x}=(r_{S}F_{S}/\theta _{y})[\tanh \theta _{S}\tanh \theta _{SN}+\tanh
\theta _{y}(\tanh \theta _{S}+\tanh \theta _{SN})]/{\cal D}_{x}  \label{F_1}
\end{equation}

\begin{equation}
F_{y}=(r_{S}F_{S}/\theta _{y})[\tanh \theta _{SN}+\tanh \theta _{y}(1+\tanh
\theta _{S}\tanh \theta _{SN})]/{\cal D}_{y}  \label{F_2}
\end{equation}
where ${\cal D}_{x}=\tanh \theta _{y}\tanh \theta _{S}\tanh \theta
_{SN}+\tanh \theta _{S}+\tanh \theta _{SN};$

${\cal D}_{y}=1+\tanh \theta _{SN}(\tanh \theta _{y}+\tanh \theta _{S});$ $%
r_{S}=R_{1}/R_{b}$, $R_{1}=L_{1}/\sigma ,$ $\theta _{y}=\kappa L_{2},$ $%
\theta _{S}=\kappa L_{1},$ $\theta _{SN}=\kappa L_{SN},$ $%
L_{SN}=L_{N}-L_{1}. $

The exponential smallness of the critical current $I_{c}$ at high
temperatures $T>D/L^{2}$ can be easily verified if one considers a simple
case of equal distances $L_{1}=L_{2}=L_{SN}\equiv L$, that is, $\theta
_{S}=\theta _{y}=\theta _{SN}\equiv \theta $. In this case we get for the
difference $(F_{y}-F_{x})$

\begin{equation}
F_{y}-F_{x}=(r_{S}F_{S}/\theta )\tanh \theta \frac{(1-\tanh ^{2}\theta )^{2}%
}{(2+\tanh ^{2}\theta )(1+2\tanh ^{2}\theta )}  \label{F_2-F_1}
\end{equation}

It is seen from Eq.(\ref{F_2-F_1}) that at $\theta >>1$ we have $%
(F_{y}-F_{x})\propto (1-\tanh ^{2}\theta )^{2}\propto (1/\theta )\exp (-4L%
\sqrt{2\pi T/D}),$ that is, this difference is small if $2\omega \approx
2\pi T>>D/(2L)^{2}$. Therefore the critical Josephson current also is
exponentially small (see Eq.(\ref{I_c})). On the other hand, neither $V_{+}$
nor $V_{-}$ can be represented as a sum over the Matsubara frequencies.
Thus, one can not claim that, for example, $V_{+}$ should be exponentially
small if the Josephson coupling is very weak.

\section{\protect\bigskip Results and discussion}

We can rewrite the expressions for $V_{\pm }$\ in the form

\begin{equation}
eV_{+}/\delta T\cong -(L_{1}/L_{N})<(\epsilon \beta )F_{-}^{\prime
}F_{S}^{\prime \prime }>_{\epsilon }\sin \varphi /<2\nu _{S}+[F_{+}^{\prime
\prime }+F_{-}^{\prime \prime }\cos \varphi ]F_{S}^{\prime \prime
}>_{\epsilon };  \label{V_pA}
\end{equation}

\begin{equation}
2eV_{-}/\delta T\cong r_{S}<F_{-}^{\prime \prime }F_{S}^{\prime }>_{\epsilon
}\sin \varphi .  \label{V_mA}
\end{equation}
where $r_{S}=L_{1}/R_{b}\sigma $,\ $F_{\pm }^{\prime }=%
%TCIMACRO{\func{Re}}%
%BeginExpansion
\mathop{\rm Re}%
%EndExpansion
(F_{y}\pm F_{x}),$\ $F_{S}^{\prime \prime }=%
%TCIMACRO{\func{Im}}%
%BeginExpansion
\mathop{\rm Im}%
%EndExpansion
F_{S}$\ etc. The angle brackets mean the averaging over energies: $%
<...>=\int d\epsilon (...)f_{eq}^{\prime }$, and the functions $F_{x,y}$\
are given by Eqs.(\ref{F_1},\ref{F_2}). Using these formulae, one can
numerically calculate the thermoemf as a function of temperature for
different values of parameters.

In Fig.1 and Fig.2 we plot the dependence of the normalized voltage $%
v_{+}\equiv eV_{+}/\delta T$ and the normalized critical current $%
i_{c}\equiv I_{c}(T)/I_{c}(0)$ as a function of the normalized temperature $%
T/\epsilon _{S}$. These curves are presented for $\varphi =\pi /2$, where $%
v_{+}$ reaches a maximum, and different values of the ratio $\Delta
/\epsilon _{S}$. We choose the value of $r_{S}$ equal to: $r_{S}=0.5$ (note
that the curves for smaller $r_{S}$, $r_{S}$ $\leq 0.3,$ are similar to
those plotted in Fig.1 and 2). We set $\gamma =0.1\epsilon _{S}$\ for the
data in Fig.1 and $\gamma =0.03\epsilon _{S}$\ for the data in Fig.2. Note
that the damping parameter $\gamma $\ may essentially exceed the
corresponding value in bulk superconductors because the PE leads to an
additional damping of the order of $\alpha r_{S}\epsilon_{S} $, where $%
\alpha $\ is a parameter which depends on geometry of the S and N films. We
see that in both figures the critical current decays to zero smoothly and
fast (faster than exponentially) when the temperature $T$ increases, but the
magnitude of $v_{+}$ decreases not so fast. Close to the critical
temperature $T_{c}$ the critical current is very small in comparison with $%
I_{c}(0)$, whereas the normalized thermoemf is not so small.

\begin{figure}[tbp]
\epsfysize= 8cm \vspace{0.2cm}
\centerline{\epsfbox{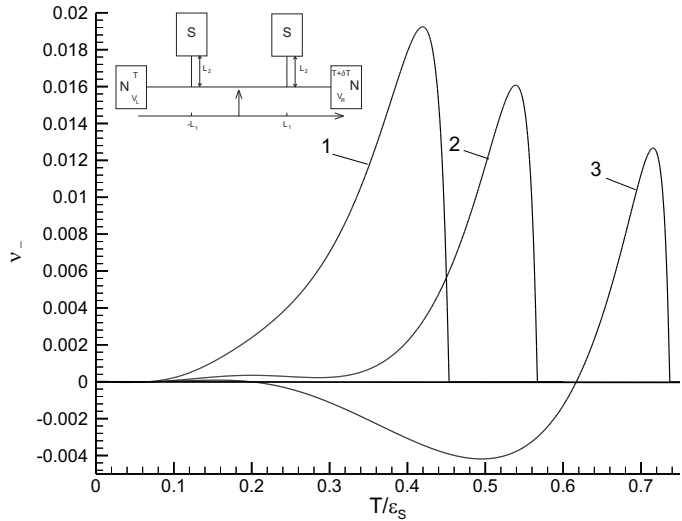}} \vspace{0.2cm}
\large{\caption{Temperature dependence of the normalized thermoelectric voltage between
the normal reservoirs $v_{-}=e(V_{r}-V_{l})/2\delta T$ for different values
of the ratio $\Delta /\epsilon _{S}$ and $r_{S}=0.5$ (the values of other parameters are the same
as in Fig.2). Inset: the structure
under consideration; the normal wire connects two normal reservoirs kept at
temperatures $T$ (left reservoir) and $T+\delta T$ (right reservoir) and two
superconductors with the phases $\varphi /2$ and $-\varphi /2.$ The distance
between the normal reservoirs is: $L_{N}=L_{1}+L_{SN}.$ The electric
potential of the superconductors is set to zero.}}

\end{figure}

Although the value of $v_{-}\equiv eV_{-}/\delta T$ is much smaller than $%
v_{+}$, it depends on temperature quite differently being very small at low $%
T$ and having a maximum at a temperature near $T_{c}.$ We represent the
temperature dependence of $v_{-}$ in Fig.3. As we mentioned before, the
voltage $eV_{-}$ is very small at low temperatures $T$ because the function $%
g_{z-}(\epsilon ,L_{1})$ is zero at $|\epsilon |<\Delta $ and grows for $%
T>\Delta $. At $T=T_{c}$ this voltage as well as $V_{+}$ turns to zero.
Therefore, for some parameters the voltages $V_{r,l}=(V_{+}\pm V_{-})$ may
have two maxima (or extrema): one at low $T$, where $V_{+}$ has a maximum,
and another one at a temperature close to $T_{c},$ where $V_{-}$ has a
maximum. It is worth mentioning that a possible assymetry of the system (for
example, different distances between the crossing point and the left and
right N reservoirs) may increase the voltage $V_{-}$\ \cite{H}.

In Fig.4 and 5 we plot the temperature dependence $v_{r}=v_{+}+v_{-}$ for
the damping parameters $\gamma =0.1$ and $\gamma =0.03.$ One can see that
this dependence has two maxima. The resistance of a S/N system has a similar
temperature dependence with two maxima \cite{Shapiro}. Moreover, for $\Delta
/\epsilon _{S}\succcurlyeq 1.5$ the thermoemf drops to zero at a certain
temperature, remains very small at larger $T$ and increases again at
temperatures close to $T_{c}$. At $T=T_{c}$ the thermoemf drops to zero. A
similar dependence of the thermoemf $V_{r}$ as a function of the magnitude
of the dc heater current was observed in experiment \cite{Petrashov}.

Note that the voltage $V_{l}=V_{+}-V_{-}$\ of the left cool N reservoir (see
Fig.6) has the second maximum near $T_{c}$\ the sign of which is opposite to
the sign of the first maximum . Such a behaviour was observed in experiment  
\cite{Sosnin}.

Obviously the average temperature in the N wire (or film) is proportional to
the magnitude of the heater current. Thus one can say about a qualitative
agreement between theory and experiment. It is difficult to carry out more
precise comparison because the PE in the experiment seems to be strong.
Perhaps, the strong PE is the reason for a high value of the second maximum
of $V_{r}$ near $T_{c}$ (in the experiment, it is comparable with the value
of the first maximum at low $T$) because it is proportional to the parameter 
$r_{S}.$

\begin{figure}[tbp]
\epsfysize= 8cm \vspace{0.2cm}
\centerline{\epsfbox{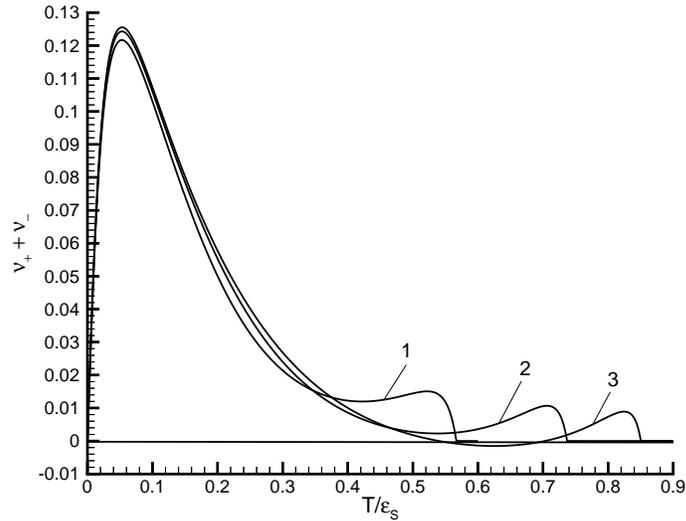}} \vspace{0.2cm}
\large{\caption{Temperature dependence of the thermoelectric voltage $%
v_{r}=(v_{+}+v_{-})=eV_{r}/\delta T$ for different ratio  $\Delta /\epsilon
_{S}$ and $r_{S}=0.5$ (the values of other parameters are the same
as in Fig.1).}}
\end{figure}

\begin{figure}[tbp]
\epsfysize= 8cm \vspace{0.2cm}
\centerline{\epsfbox{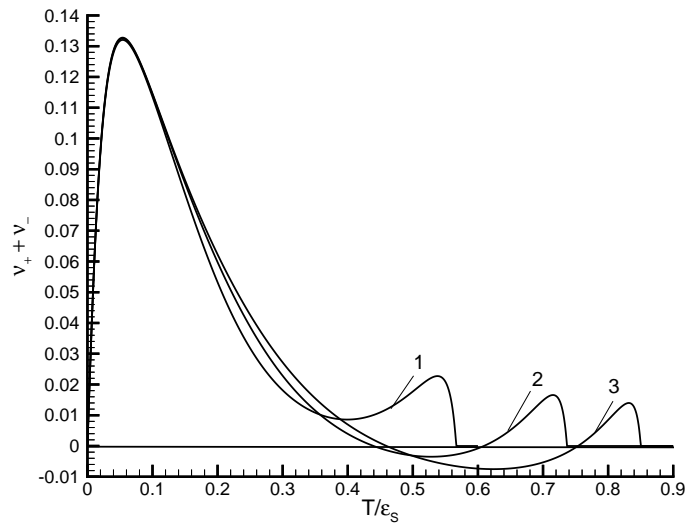}} \vspace{0.2cm}
\large{\caption{The same dependencies as in Fig.4 for
$\gamma =0.03\epsilon _{S}$.}}

\end{figure}

We assumed that the parameter $r_{S}$ is small; however in the case of a
strong PE this parameter should be taken of the order of 1 because the
condensate functions $F^{R(A)}$ in the N film saturate with increasing $r_{S}
$. Finally we give some estimations for parameters close to the experimental
ones \cite{Petrashov}: for $L_{1}=0.2\mu $ and $D=150cm^{2}/s$, one has $%
\epsilon _{S}\approx 2.8K.$ This value of $\epsilon _{S}$ is a little larger
than $\Delta \approx 2.2K$ ($\Delta \approx 0.8\epsilon _{S}$). The
estimations of the thermoemf give the values $eV_{th}/\delta T\approx
6-10\mu V/K$. These values are comparable with the voltage $V_{th}$ observed
in Ref. \cite{Chandrasekhar98}, but much larger than $V_{th}$ measured in
Refs. \cite{Chandrasekhar02,Petrashov} {\bf (}note that in Refs.\cite
{Chandrasekhar98,Chandrasekhar02} the voltage $V_{-},$\ but not $V_{+}$\ was
measured; in the case $r_{S}\sim 1$\ the magnitudes of both voltages are
comparable). Perhaps the reason for this discrepancy is the energy
relaxation processes in the N wire which were neglected in our calculations.
Indeed, in order to neglect the inelastic scattering the condition $\tau
_{in}^{-1}<<r_{S}(D/L^{2})$\ should be fulfilled, where $\tau _{in}^{-1}$\
is the inelastic scattering rate. Otherwise in the system arises a strong
depairing which suppresses the PE and destroys the phase coherent effects
discussed above. The more exact comparision with experimental data is not
possible at this stage; for example, a thermoemf symmetric with respect to
the phase difference $\varphi $ was observed in experiments. The origin of
this part of thermoemf stll is unclear.

\begin{figure}[tbp]
\epsfysize= 8cm \vspace{0.2cm}
\centerline{\epsfbox{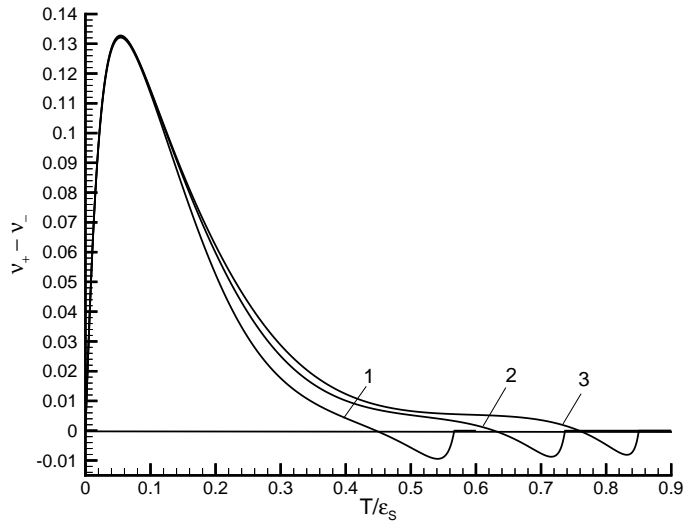}} \vspace{0.2cm}
\large{\caption{Temperature dependence of the thermoelectric voltage $%
v_{l}=(v_{+}-v_{-})=eV_{r}/\delta T$ for different ratio  $\Delta /\epsilon
_{S}$, for $r_{S}=0.5$ (the values of other parameters are the same
as in Fig.2).}}
\end{figure}

\section{\protect\bigskip Conclusion}

\bigskip

In conclusion, we show that the thermoemf arising in the four-terminal S/N
structure (see inset in Fig.2) is measurable even in the case of \ a very
weak Josephson coupling. One can say about a long-range thermoemf. The
thermoemf $V_{l,r}$ (or thermopower $V_{l,r}/\delta T$) depends on $T$ in a
non-monotonic way and may have two extrema: one at low temperatures and
another at temperatures close to $T_{c}$. At some values of parameters and
intermediate temperatures the thermoelectric voltage is negligible. This
behavior qualitatively agrees with the experimental observations \cite
{Petrashov}.

We would like to thank SFB 491 for a financial support. We are grateful to
V. Chandrasekhar, V.Petrashov and I.Sosnin for valuable discussions.


\begin{references}
\bibitem{Foot}  Note that the density-of-states in the superconductors $%
\upsilon _{S}$ is dropped in the expression for $V_{+}$ in Ref. \cite{KPV}
because at low $T$ it is zero. For $T>\Delta $ it should be taken into
account.
\end{references}
\end{document}